\documentclass[conference]{IEEEtran}
\usepackage{graphicx}
\usepackage{amsmath}
 
\usepackage{booktabs}
\usepackage{caption}
\usepackage{tikz}
\usepackage{algorithm}
\usepackage{algpseudocode}

\usetikzlibrary{positioning, shapes, arrows}

\usepackage{hyperref}

\title{sVIRGO: A Scalable Virtual Tree Hierarchical Framework for Distributed Systems}

\author{
    \IEEEauthorblockN{Lican Huang}
    \IEEEauthorblockA{Hangzhou Domain Zones Technology Co., Ltd. \\ Email: lican.huang.hz@gmail.com}
}

\begin{document}

\maketitle

 \begin{abstract}
 
 We propose sVIRGO, a scalable virtual tree hierarchical framework for large-scale distributed systems. sVIRGO constructs virtual hierarchical trees directly on physical nodes, allowing each node to assume multiple hierarchical roles without overlay networks. The hierarchy preserves locality and is organized into configurable layers within regions. Coordination across thousands of regions is achieved via virtual upper-layer roles dynamically mapped onto nodes up to the top layer.

Each region maintains multiple active coordinators that monitor local health and perform dynamic re-selection if failures occur. Temporary drops below the minimum threshold do not compromise coordination, ensuring near-zero recovery latency, bounded communication overhead, and exponentially reduced failure probability while maintaining safety, liveness, and robustness under mobile, interference-prone, or adversarial conditions.

Communication is decoupled from the hierarchy and may use multi-frequency wireless links. Two message hop strategies are supported: (i) with long-distance infrastructure-assisted channels, coordinators exploit the virtual tree to minimize hops; (ii) without such channels, messages propagate via adjacent regions.  
 
 sVIRGO also  supports  Layer-Scoped Command Execution. Commands and coordination actions are executed within the scope of each hierarchical layer, enabling efficient local and regional decision-making while limiting unnecessary global propagation.
 
\end{abstract}


 \section{Introduction}
Large-scale distributed systems, \textbf{excluding overlay-based architectures}, increasingly underpin critical applications such as autonomous drone swarms, large-scale Internet of Things (IoT) deployments,  and cyber-physical systems. As the number of participating nodes grows to the order of millions, these systems face fundamental challenges in coordination, communication efficiency, fault tolerance, and scalability. Traditional flat mesh, or ad-hoc designs incur excessive communication overhead, require costly global state maintenance, and often become fragile under node failures, mobility, or adversarial disruptions.

Hierarchical architectures have long been explored as a means to improve scalability. However, conventional hierarchical approaches typically rely on rigid structures, fixed roles, or overlay networks that introduce additional complexity, maintenance overhead, and performance bottlenecks. Moreover, many existing designs assume a static mapping between physical nodes and hierarchical positions, limiting their adaptability to dynamic environments and large-scale disruptions.

To address these limitations, we propose \textbf{sVIRGO}, a scalable virtual tree hierarchical framework that operates \emph{directly on physical nodes}, without relying on overlay networks or dedicated infrastructure. In sVIRGO, hierarchical roles are virtualized and dynamically mapped onto physical nodes, allowing each node to concurrently assume multiple roles across different layers of the hierarchy. This design enables flexible role assignment, locality-aware coordination, and efficient communication while maintaining scalability at extreme system sizes.

Starting from local regions, sVIRGO organizes nodes into a configurable multi-layer virtual hierarchy that preserves physical locality and minimizes communication overhead. At larger scales, coordination across thousands of regions is achieved through multiple virtual upper-layer roles dynamically mapped onto physical nodes within regions, extending up to the top layer of the system. Importantly, even top-layer global roles remain purely virtual and are not bound to special or centralized physical nodes, eliminating single points of failure and improving system robustness.

To further enhance resilience, sVIRGO incorporates \textbf{regional multi-coordinator redundancy} with fully local monitoring and dynamic role re-selection. Each region maintains multiple active \textbf{coordinators}, ensuring that transient failures or mobility-induced disruptions do not compromise regional coordination. Even when the number of active coordinators temporarily falls below the minimum threshold, local re-selection restores operational capacity with near-zero expected recovery latency and bounded overhead.

Communication in sVIRGO is decoupled from the hierarchical structure itself, allowing the system to adapt to diverse deployment conditions. sVIRGO supports hybrid communication strategies: when long-distance infrastructure-assisted links (e.g., 5G/6G or satellites) are available, coordinators leverage the virtual hierarchy to minimize message hops; when such infrastructure is unavailable, messages are forwarded through adjacent regions using locality-preserving multi-hop paths. This flexibility enables sVIRGO to operate efficiently in infrastructure-rich, infrastructure-free, or adversarial environments while mitigating interference, jamming, and targeted attacks.

\subsection{Contributions}
This paper makes the following key contributions:
\begin{enumerate}
             
 \item  Scalable Virtual Hierarchical Framework: We propose \textbf{sVIRGO}, a virtual tree hierarchical architecture that allows nodes to assume multiple roles across layers without relying on overlay networks, enabling scalable coordination for million-node distributed systems. We present a scalable region-to-global coordination architecture in which multiple virtual upper-layer roles, spanning from local regions up to the top layer, are dynamically mapped onto physical nodes, allowing each node to concurrently assume multiple hierarchical roles.
 
 \item Regional Active-Coordinator Re-Selection for Resilient sVIRGO.
We propose a novel \textbf{regional redundancy strategy} in sVIRGO, where each \textbf{region} maintains multiple active \textbf{coordinators}. Monitoring and \textbf{re-selection} are performed entirely \textbf{locally}, ensuring that even if the number of active \textbf{coordinators} temporarily drops below the minimum threshold $\mathbf{T_{\min}}$, the \textbf{region} automatically restores its operational threshold. This approach guarantees \textbf{near-zero expected recovery latency}, \textbf{bounded communication overhead}, and \textbf{exponentially reduced failure probability}, while maintaining \textbf{safety, liveness, and robustness} even in mobile, interference-prone, or adversarial environments. To our knowledge, this is the first design that combines \textbf{multi-coordinator redundancy}, \textbf{local re-selection}, and \textbf{provable correctness guarantees} in a hierarchical virtual overlay network.

 \item Layer-Scoped Command Execution: Commands and coordination actions are executed within the scope of each hierarchical layer, enabling efficient local and regional decision-making while limiting unnecessary global propagation.

 \item Infrastructure-Optional Communication: sVIRGO supports hybrid communication, leveraging infrastructure-assisted links (e.g., 5G/6G or satellites) when available, or relying on multi-hop forwarding over worker-level ad-hoc networks when infrastructure is absent.
 
\end{enumerate}       

 \section{Related Work}

\subsection{Non-Hierarchical Architectures}
Flat architectures, including mesh and ad hoc networks, provide fully decentralized communication among nodes, with no predefined roles or layered structure. In multi-UAV swarms, wireless sensor networks, and mobile ad hoc networks (MANETs), routing protocols such as  Optimized Link State Routing (OLSR) \cite{clausen2003olsr}, Ad hoc On-Demand Distance Vector (AODV) \cite{perkins1999aodv},    Dynamic Source Routing (DSR) \cite{johnson1996dsr}
 and Better Approach To Mobile Adhoc Networking (BATMAN) \cite{neumann2008batman}   have been widely adopted for distributed coordination and data exchange \cite{bekmezci2013flying}. These protocols enable multi-hop communication through route discovery, periodic link-state dissemination, or opportunistic forwarding, allowing nodes to dynamically adapt to mobility and topology changes.

Flat designs offer robustness at small to medium scales, as the absence of centralized control avoids single points of failure. However, scalability becomes a fundamental limitation as the number of nodes increases. Proactive protocols, such as OLSR, incur substantial control overhead due to periodic topology broadcasts, while reactive protocols, such as AODV and DSR, suffer from increasing route discovery latency and flooding overhead as network size grows. Maintaining routing tables, neighbor states, and link-quality information becomes increasingly expensive at large scales.

In large UAV swarms, flat architectures also face severe communication contention and interference, particularly in shared wireless spectra. As node density increases, broadcast storms, packet collisions, and congestion significantly degrade throughput and reliability. Moreover, global coordination tasks, such as swarm-wide state aggregation, task assignment, or synchronized maneuvering, require extensive multi-hop communication, leading to high latency and poor responsiveness.

From a security and resilience perspective, flat ad hoc networks are vulnerable to targeted attacks, jamming, and malicious nodes. Disrupting a subset of highly connected nodes or injecting false routing information can propagate widely due to the lack of structural containment. These limitations motivate the adoption of hierarchical and structured architectures to improve scalability, efficiency, and robustness in ultra-large-scale distributed systems.

\subsection{Hierarchical Architectures}

Hierarchical  architectures introduce structured organization by grouping nodes into clusters, layers, or roles, enabling scalable coordination and control. Existing hierarchical approaches can be broadly categorized into static hierarchical swarms and dynamic hierarchical swarms, depending on whether the hierarchy is predefined or adaptively formed.

\subsubsection{Static Hierarchical Architectures} 

In static hierarchical architectures, the roles of nodes (e.g., leaders, cluster heads, or relay nodes) are predefined before deployment and remain fixed throughout operation. Such designs are common in military UAV formations, satellite constellations, and industrial monitoring systems, where mission parameters and communication topology are known in advance \cite{hayat2016survey}.

Protocols such as Cluster-Based Routing Protocol (CBRP) \cite{jiang1999cbrp} and Hierarchical OLSR (HOLSR) \cite{yingge2005} organize nodes into clusters with designated leaders responsible for intra- and inter-cluster communication. This structure significantly reduces routing overhead and improves scalability compared to flat networks. However, static hierarchies suffer from limited adaptability: failures or mobility of key nodes can lead to performance degradation or network partitioning. Moreover, fixed leadership introduces single points of failure and makes the system vulnerable to targeted attacks or overload conditions.

\subsubsection{Dynamic Hierarchical Architectures}

Dynamic hierarchical architectures aim to overcome these limitations by allowing the hierarchy to evolve in response to network conditions, node mobility, and task requirements. Leadership roles and cluster memberships are assigned adaptively based on metrics such as connectivity, energy level, mobility patterns, or computational capacity \cite{gupta2016survey}.

Representative approaches include  Low-Energy Adaptive Clustering Hierarchy (LEACH) \cite{palan2017} and   intelligent bio-inspired multi-objective and scalable UAV-assisted clustering algorithm in flying ad hoc networks  \cite{Aslam2026}. In UAV swarms, dynamic clustering enables efficient local coordination while preserving global scalability. By rotating leadership roles and redistributing responsibilities, these systems improve robustness against node failures and adversarial disruptions.

Nevertheless, dynamic hierarchies introduce additional complexity and coordination overhead. Frequent re-clustering may lead to transient instability, increased signaling cost, and synchronization challenges. Ensuring consistency across hierarchical levels while maintaining low latency remains an open problem, particularly in ultra-large-scale swarms with heterogeneous agents and rapidly changing environments.

\subsection{sVIRGO }

When scaled to millions of nodes, existing distributed system architectures—including flat networks and static or dynamic hierarchical designs—face fundamental scalability limits. Flat mesh and ad hoc networks incur excessive routing and signaling overhead as network diameter grows. Static hierarchies lack adaptability and are vulnerable to failures, while dynamic hierarchies, despite re-clustering and leader re-election, introduce significant coordination overhead and instability. Consequently, maintaining global consistency and robustness becomes increasingly difficult under failures and mobility.

sVIRGO builds upon hierarchical design principles while introducing 
\emph{non-overlay virtual hierarchies}, multi-role physical nodes, and strict 
locality preservation. Its configurable multi-layer virtual structure allows nodes 
to participate concurrently in multiple hierarchical roles.  Dynamic role 
re-selection provides robustness against node failures and large-scale disruptions, 
and communication can leverage variable multi-frequency wireless links to mitigate 
interference, jamming, and attacks. This design allows sVIRGO to easily scale to 
millions of nodes, as illustrated in simulations of million-drone swarms spanning 
thousands of local regions, outperforming flat and static hierarchical architectures.

\section{sVIRGO Hierarchical Network Architecture}

\subsection{Overview}
sVIRGO organizes nodes into \textbf{thousands of local regions},  which  form each a  virtual hierarchical structure for scalable coordination. Only the bottom layer consists of physical nodes (workers), while all upper layers are virtual roles dynamically mapped onto these physical nodes. The number of virtual layers is configurable; in the example below, we illustrate a five-layer hierarchy.   

\subsection{Layer Definitions}

\textbf{Layer 1 – Worker Nodes (Physical):}  
The base layer exists all workers in the system. Workers are responsible for sensing, data collection, and task execution.

\textbf{Layer 2 – Cluster Leaders (Virtual):}  
Selected from worker nodes, each cluster leader coordinates its worker cluster. Redundancy is introduced to improve reliability and ensure continuous operation.

\textbf{Layer 3 – Regional Hubs (Virtual):}  
Regional hubs manage multiple clusters. By default, they are selected from cluster leaders, but in configurable deployments, they can also be selected directly from workers. Redundancy ensures fault tolerance and operational continuity.

\textbf{Layer 4 – Local Global Command Nodes  (Virtual):}  
These nodes oversee multiple regional hubs . They can be selected from regional hubs, cluster leaders, or directly from workers depending on deployment needs. Redundancy is applied to maintain resilience and enable continuous swarm-wide coordination.

\textbf{Layer 5 – Top-Level Global Roles (Virtual):}  
These nodes  coordinate with other Layer 4 nodes across regions. They can be selected from  Local Global Command Nodes. Redundancy is applied to maintain resilience and enable continuous swarm-wide coordination.

\textbf{Key Features:}  
\begin{itemize}
    \item Only Layer 1 consists of physical nodes; all higher layers are virtual roles dynamically mapped onto these nodes.
    \item Each physical node may concurrently assume multiple virtual roles.
    \item The hierarchy is locality-preserving and scalable to millions of nodes across thousands of regions.
    \item Redundancy in virtual layers ensures robustness against node failures, mobility, and attacks.
    \item Flexible node selection rules allow configurable deployment strategies for regional hubs and top-level nodes.
\end{itemize}
 
 \begin{figure*}[h]
\centering
\begin{tikzpicture}[
    node distance=1cm and 1cm,
    worker/.style={circle, draw=blue!70, fill=blue!20, minimum size=6mm},
    leader/.style={circle, draw=red!70, fill=red!20, minimum size=8mm},
    hub/.style={circle, draw=green!70, fill=green!20, minimum size=10mm},
    global/.style={circle, draw=orange!70, fill=orange!20, minimum size=12mm},
    swarmglobal/.style={circle, draw=red!80, fill=red!30, minimum size=12mm},
    every edge/.style={draw, -latex}
]

\node[worker] (w1) {W1};
\node[worker, right=of w1] (w2) {W2};

\node[worker, right=of w2] (w3) {W3};
\node[worker, right=of w3] (w4) {W4};

\node[worker, right=of w4] (w5) {W5};
\node[worker, right=of w5] (w6) {W6};

\node[worker, right=of w6] (w7) {W7};
\node[worker, right=of w7] (w8) {W8};

\node[leader, above=of w2] (cl1) {CL1/W1};
\node[leader, above=of w3] (cl2) {CL2/W4};
\node[leader, above=of w6](cl3) {CL3/W5};
\node[leader, above=of w7] (cl4) {CL4/W8};


\draw[dotted] (w1) -- (cl1);
\draw [dotted](w2) -- (cl1);

\draw[dotted] (w3) -- (cl2);
\draw[dotted] (w4) -- (cl2);

\draw [dotted] (w5) -- (cl3);
\draw [dotted] (w6) -- (cl3);

\draw [dotted] (w7) -- (cl4);
\draw [dotted] (w8) -- (cl4);

 
\node[hub, above=of cl2, xshift=-0.5cm] (rh1) {RH1/W1};
\node[hub, above=of cl3, xshift=0.5cm] (rh2) {RH2/W8};

\draw [dotted] (cl1) -- (rh1);
\draw [dotted] (cl2) -- (rh1);
\draw [dotted] (cl3) -- (rh2);
\draw [dotted] (cl4) -- (rh2);

\node[global, above=of rh1, xshift= 1cm] (lgc1) {LG1/W1};
\node[global, above=of rh2, xshift=-1cm] (lgc2) {LG2/W8};

\draw [dotted] (rh1) -- (lgc1);
\draw [dotted] (rh2) -- (lgc2);

\node[swarmglobal, above=of lgc1, xshift= 2cm] (gc1) {GC1/W8};
 
\draw [dotted] (lgc1) -- (gc1);
\draw [dotted] (lgc2) -- (gc1);
\ 
\node[left=0.5cm of w1] (l1) {Layer 1: Workers};
\node[left=0.5cm of cl1] (l2) {Layer 2: Cluster Leaders};
\node[left=0.5cm of rh1] (l3) {Layer 3: Regional Hubs};
\node[left=0.5cm of lgc1] (l4) {Layer 4: Local Global Command};
\node[left=0.5cm of gc1] (l5) {Layer 5: Global Command};
\end{tikzpicture}
 \caption{sVIRGO 5-layer virtual hierarchy within a local region. Only workers are physical nodes; all upper layers are virtual roles dynamically mapped onto these nodes. LG2/W8 also takes top-level global coordinator role.}

\label{fig:svrigo_16workers}
\end{figure*}
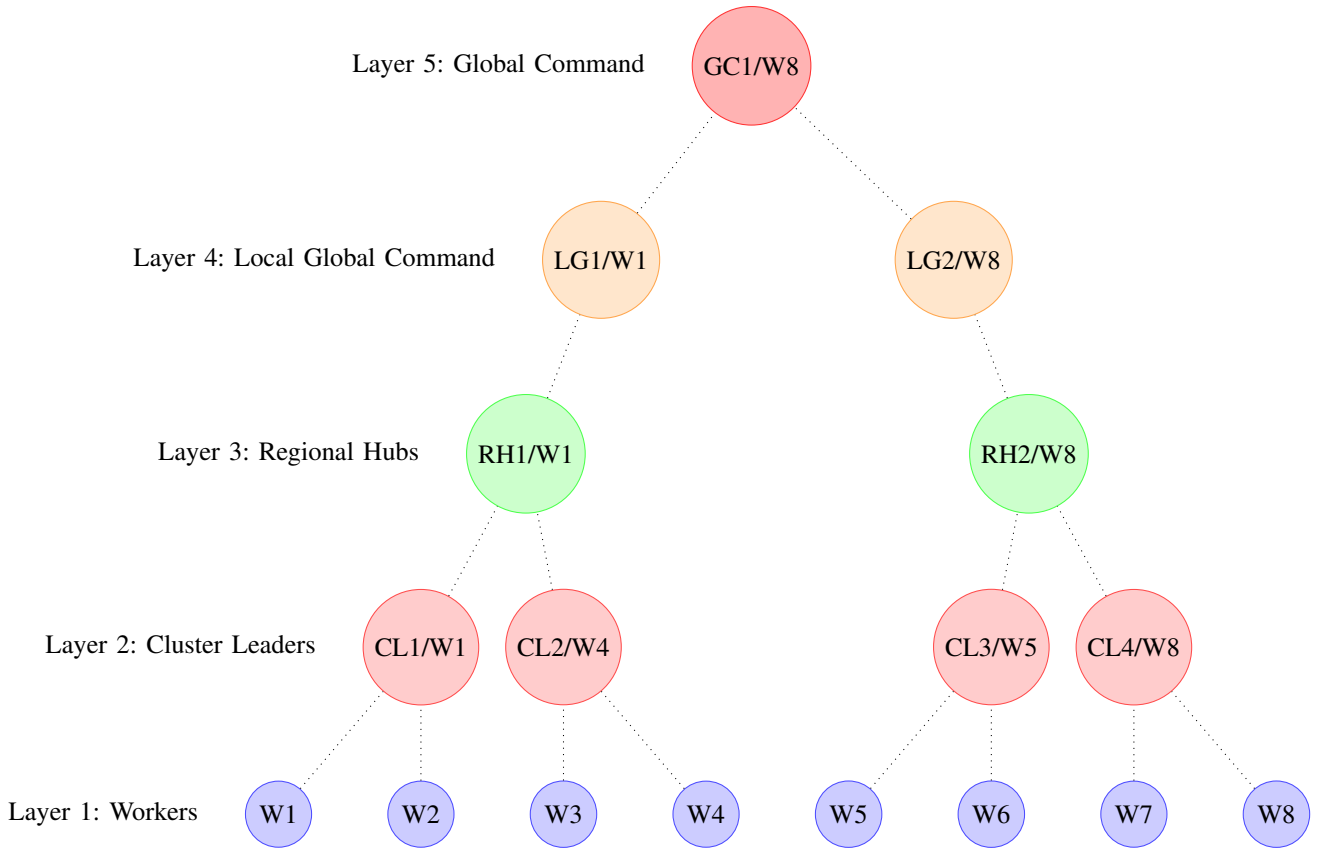

 \subsection{sVIRGO Communications}

sVIRGO is designed to support ultra-large-scale swarms comprising up to millions of agents distributed over wide geographic areas. At this scale, purely flat ad-hoc communication becomes impractical due to excessive routing overhead, long multi-hop paths, and instability induced by dynamic topology changes. Conversely, fully centralized or infrastructure-dependent communication models undermine scalability, robustness, and swarm autonomy.

At the physical layer, worker nodes can communicate using various protocols. To balance scalability, resilience, and autonomy, sVIRGO usually adopts a hybrid communication architecture. At the physical layer, worker nodes communicate using flat, short-range ad-hoc links within local neighborhoods or clusters. This communication model enables low-latency local coordination and sensing while avoiding the need for maintaining global routing state.

For long-distance and cross-region coordination, sVIRGO constructs a hierarchical virtual communication overlay composed of dynamically assigned virtual roles, including cluster leaders, regional hubs, and local global command nodes. These virtual nodes aggregate, summarize, and relay information across hierarchy levels, enabling scalable global-wide coordination without requiring direct long-range communication links among individual workers.

When available, infrastructure-assisted communication technologies—such as 5G/6G cellular networks or satellite links—may be leveraged by higher-level virtual nodes to reduce latency and increase bandwidth for inter-region communication. In the absence of such infrastructure, higher-layer virtual nodes do not rely on direct long-range links. Instead, inter-region communication is realized through multi-hop forwarding across neighboring regions, where messages propagate via flat ad-hoc links among worker nodes at regional boundaries.

Specifically, coordination messages generated by a virtual node are first disseminated within its own region and forwarded by boundary worker nodes to adjacent regions using local ad-hoc connectivity. These messages are then aggregated and relayed by corresponding virtual nodes in neighboring regions. This process repeats across regions until the intended regional or global coordinator is reached. Through this region-by-region hop propagation, upper-layer virtual roles achieve effective global coordination while relying exclusively on short-range physical communications.

By decoupling physical connectivity from virtual coordination roles, sVIRGO enables scalable, resilient, and infrastructure-optional communication, making it well suited for wide-area, million-scale swarm deployments.

\subsection{Dynamic Role Switching}

sVIRGO supports dynamic role switching to adapt to node mobility, failures, and workload imbalance. Nodes may change roles within the same hierarchy layer to maintain local connectivity and performance. Higher-layer virtual roles are dynamically re-elected from the immediate lower layer, ensuring that the hierarchical constraints of the virtual tree are preserved at all times.

Role transitions are triggered by local observations, such as energy depletion, link degradation, or leadership overload, and are coordinated through neighborhood-level consensus. Since all upper-layer roles are virtual and mapped onto physical worker nodes, role switching does not require physical redeployment and incurs minimal control overhead. This dynamic reassignment mechanism enables sVIRGO to maintain scalability, robustness, and continuity of coordination under highly dynamic swarm conditions.

\subsection{Message Hops without long distance communication}

In sVIRGO, command dissemination operates \emph{without direct assistant-to-assistant communication}. Instead, message propagation is realized through a \emph{hierarchical cluster--worker system}, in which cluster leaders coordinate routing decisions and workers act as stateless forwarders. Clusters are organized into regions, and clusters within the same region are considered \emph{distance-adjacent} and logically coordinated by a region-level layer. Neighboring regions are further grouped under hub and global scopes, forming a multi-layer hierarchy. 

Message forwarding exploits this structure by prioritizing clusters that are topologically closer in the hierarchy—clusters in the same region are treated as nearest neighbors, followed by clusters within the same hub, and finally those at the global scope. This hierarchical neighbor-distance model enables efficient, loop-free message delivery across large-scale deployments while supporting multi-goal execution and controlled propagation termination.

\subsubsection{Message Structure}

Commands are propagated as cluster messages:

 \[
\begin{aligned}
M = \langle\;&
\texttt{goalClusterIDs},\\
&\texttt{targetWorkerIDs},\\
&\texttt{visitedClusterIDs},\\
&\texttt{executedClusterIDs},\\
&\texttt{hopCount},\\
&\texttt{originalSource}, \\
&\texttt{lastSentClusterID},\\
&\texttt{forwardFlag}  
\;\rangle
\end{aligned}
\]

\begin{itemize}
    \item \textbf{goalClusterIDs}: the set of valid destination clusters derived from the command scope (region, hub, local-global, or global).
    \item \textbf{targetWorkerIDs}: optional workers within a cluster that should execute the command. If empty, execution is handled by cluster-level policy.
    \item \textbf{visitedClusterIDs}: clusters that have already \emph{received} the message, used for loop prevention and suppression.
    \item \textbf{executedClusterIDs}: clusters that have \emph{completed execution}; forwarding only considers clusters not yet executed.
    \item \textbf{hopCount}: number of cluster-level hops traversed; used for propagation control and diagnostics.
 
  \item \textbf {originalSource}:	  cluster ID that first created the message
\item \textbf {lastSentClusterID}:	Cluster ID from which the message was most recently forwarded
\item \textbf { forwardFlag}:  Boolean: workers authorized to broadcast
\end{itemize}
\subsubsection{Hierarchical Distance and Deferred Forwarding}

Each cluster leader estimates the hierarchical distance to all unexecuted goal clusters:
\begin{itemize}
    \item Same cluster: distance 0
    \item Same region: distance 1
    \item Same hub: distance 2
    \item Local-global domain: distance 3
    \item Global scope: distance 4
\end{itemize}

A deferred forwarding delay is computed as:
\[
\text{delay} = \alpha \cdot \text{distance} + \beta \cdot \text{localLoad} + \text{rand}(0, \epsilon)
\]
which prioritizes closer clusters and reduces redundant messages.

\subsubsection{Algorithm 1: Worker-Mediated Forwarding}

Algorithm~1 describes the behavior of \textbf{workers} in sVIRGO during message dissemination. Workers are primarily passive execution entities: they execute commands only when explicitly targeted and do not initiate forwarding on their own. Upon first receiving a message from a neighboring cluster, a worker reports the message to its local \textbf{cluster leader}, enabling leader-driven coordination and routing decisions. Worker-level message propagation occurs \emph{only} when explicitly authorized by the leader via a forwarding flag, ensuring controlled dissemination, loop avoidance, and bounded communication overhead. This design separates execution from routing, prevents uncontrolled broadcasts, and allows leaders to enforce hierarchical and locality-aware forwarding policies.

 \begin{algorithm}[H]
\caption{Worker-Mediated Forwarding}
\begin{algorithmic}[1]
\State \textbf{upon receive($M$) at worker $W$ do}

\If{$W \in M.targetWorkerIDs$}
    \State execute command locally
\EndIf

\If{$W.clusterID \notin M.visitedClusterIDs$}
    \If{$M.lastSentClusterID \neq W.clusterID$}
        \Comment{first reception by this cluster from neighbor source}
        \State report $M$ to local cluster leader
    \EndIf
\EndIf

\If{$M.forwardFlag = \texttt{true}$}
    \If{$M.lastSentClusterID = W.clusterID$}
        \Comment{leader-authorized propagation; flat worker-layer broadcast}
        \State broadcast $M$ to all reachable workers
    \EndIf
\EndIf

\end{algorithmic}
\end{algorithm}

 \subsubsection{Algorithm 2: Cluster Leader Deferred Routing with Execution Tracking}
 
Algorithm~2 presents a \emph{deferred routing} strategy executed by a cluster \textbf{leader} when immediate hierarchical forwarding is either unavailable or undesirable. Instead of forwarding messages immediately, the leader first records local execution state and evaluates the remaining unexecuted goal clusters. If additional forwarding is required, the leader computes a delay based on hierarchical distance, local load, and randomized jitter, and only then authorizes a controlled worker-level broadcast. This mechanism prevents redundant propagation, mitigates congestion under high load, and preserves loop freedom through explici

 \begin{algorithm}[H]
\caption{Cluster Leader Deferred Routing}
\begin{algorithmic}[1]
\State \textbf{upon receive($M$) at cluster leader $C$ do}

\If{$C.clusterID \in M.visitedClusterIDs$}
    \State drop $M$
    \State \textbf{return}
\EndIf

\State $M.visitedClusterIDs \gets M.visitedClusterIDs \cup \{C.clusterID\}$

\If{$C.clusterID \in M.goalClusterIDs$}
    \State deliver $M$ to $M.targetWorkerIDs$
    \State $M.executedClusterIDs \gets M.executedClusterIDs \cup \{C.clusterID\}$
\EndIf

\State $unexecutedGoals \gets M.goalClusterIDs \setminus M.executedClusterIDs$

\If{$unexecutedGoals = \emptyset$}
    \State \textbf{return} \Comment{all goal clusters executed, stop forwarding}
\EndIf

\State $dist \gets \min_{g \in unexecutedGoals} \text{hierarchyDistance}(C,g)$
\State $delay \gets \alpha \cdot dist + \beta \cdot localLoad + rand(0,\epsilon)$
\State schedule \textbf{workerBroadcast($M$)} after $delay$

\Procedure{workerBroadcast}{$M$}
    \State $M.hopCount \gets M.hopCount + 1$
    \State $M.lastSentClusterID \gets C.clusterID$ \Comment{mark message origin for workers}
    \State $M.forwardFlag \gets \texttt{true}$ \Comment{authorize workers to broadcast}
    \State broadcast $M$ to all local workers
\EndProcedure

\end{algorithmic}
\end{algorithm}

\subsubsection{Brief  Summary of Message Hops without long distance communication }

sVIRGO achieves scalable, loop-free message dissemination in a hierarchical cluster-worker system without relying on direct assistant-to-assistant communication. The key points are:

\begin{itemize}
    \item Workers function as stateless bridges, forwarding messages and enabling cross-region propagation without maintaining complex routing tables.
    \item Cluster leaders track execution flags (\texttt{executedClusterIDs}) to record completed goal clusters, preventing redundant forwarding and ensuring messages are only delivered once per goal cluster.
    \item Deferred forwarding based on hierarchical distance prioritizes propagation: clusters closer to unexecuted goal clusters broadcast earlier, while farther clusters introduce a distance- and load-dependent delay:
    \[
    delay = \alpha \cdot dist + \beta \cdot localLoad + rand(0,\epsilon)
    \]
    where $dist$ is the minimum hierarchy distance to unexecuted goals, $localLoad$ represents current cluster activity, and $rand(0,\epsilon)$ introduces minor randomization to reduce collisions.
    \item Messages carry \texttt{forwardFlag} and \texttt{lastSentClusterID} to authorize worker-layer broadcasts and indicate the message’s origin, controlling propagation and preventing unnecessary hops.
 
    \item The combination of hierarchical leader routing and flat worker-level broadcasts reduces the possibility of redundant message hops and helps lower network congestion.

    \item This approach naturally terminates once all goal clusters have executed the command, maintaining efficient and orderly dissemination even in large-scale or cross-region deployments.
\end{itemize}

\subsection{Message Hops with long distance communication}

In sVIRGO, command dissemination is performed entirely within a \emph{hierarchical cluster-worker system} without assistants. Cluster leaders propagate messages efficiently upward to the root and downward to goal clusters. Workers only execute commands if they are explicitly targeted; they do not forward messages.

\subsubsection{Message Structure}

Messages are defined as:
 \[
M = 
\begin{aligned}
&\langle \\
&\quad \texttt{goalClusterIDs},\\
&\quad \texttt{targetWorkerIDs},\\
&\quad \texttt{visitedClusterIDs},\\
&\quad \texttt{executedClusterIDs},\\
&\quad \texttt{lastSentClusterID},\\
&\quad \texttt{hopCount} \\
&\rangle
\end{aligned}
\]

\begin{itemize}
    \item \texttt{goalClusterIDs}: clusters that must execute the command.
    \item \texttt{targetWorkerIDs}: workers within the cluster that should execute the command.
    \item \texttt{visitedClusterIDs}: clusters that have already received the message.
    \item \texttt{executedClusterIDs}: clusters that have executed the command.
    \item \texttt{lastSentClusterID}: last cluster to forward the message, used to prevent loops.
    \item \texttt{hopCount}: number of hops traversed by the message.
\end{itemize}

\subsubsection{Algorithm 3: Cluster Leader Immediate Routing with Execution Tracking}
To efficiently propagate messages within a region while avoiding redundant transmissions, each cluster leader in sVIRGO performs immediate routing with execution tracking. Upon receiving a message, a leader first checks whether its cluster has already executed the message, delivering the message to target workers if it is a goal cluster. The leader then forwards the message upward toward the root and downward along hierarchical links only to unexecuted goal clusters. This mechanism ensures loop-free dissemination, minimizes unnecessary hops, and allows precise command execution tracking across clusters.

\begin{algorithm}[H]
\caption{Cluster Leader Immediate Routing}
\begin{algorithmic}[3]
\State \textbf{upon receive($M$) at cluster leader $C$ do}

\If{$C.clusterID \in M.visitedClusterIDs$}
    \State drop $M$
    \State \textbf{return}
\EndIf

\State $M.visitedClusterIDs \gets M.visitedClusterIDs \cup \{C.clusterID\}$

\If{$C.clusterID \in M.goalClusterIDs$}
    \State deliver $M$ to $M.targetWorkerIDs$
    \State $M.executedClusterIDs \gets M.executedClusterIDs \cup \{C.clusterID\}$
\EndIf

\State $unexecutedGoals \gets M.goalClusterIDs \setminus M.executedClusterIDs$
\If{$unexecutedGoals = \emptyset$}
    \State \textbf{return} \Comment{all goal clusters executed, stop forwarding}
\EndIf

\State \textbf{forward $M$ upward to root and downward to unexecuted goal clusters via hierarchical links}
\State $M.hopCount \gets M.hopCount + 1$
\State $M.lastSentClusterID \gets C.clusterID$
\end{algorithmic}
\end{algorithm}

\subsubsection{Brief  Summary of Message Hops with long distance communication  }

\begin{itemize}
    \item sVIRGO achieves scalable, loop-free message dissemination entirely without assistants.  
    \item Workers execute commands only if targeted; they do not forward messages.  
    \item Cluster leaders track executed clusters and immediately propagate messages along hierarchical links.  
    \item Long-distance hierarchical links allow leaders to reach distant goal clusters without worker broadcasts, minimizing hops and reducing network congestion.
\end{itemize}

\subsection{Region-Scoped Redundant Active Coordinators and Resilience Analysis}

sVIRGO employs a multi-level virtual hierarchy to enable scalable coordination. However, global or multi-layer coordinator re-selection can introduce instability and cascading elections. To mitigate this, sVIRGO adopts a \textbf{region-scoped redundant active coordinator strategy}, where each region maintains a small set of active coordinators (e.g., $K$ coordinators) for local control.

Each region maintains a \textbf{minimum threshold} of $T_{\min}$ active coordinators (typically $T_{\min}=3$). When one coordinator fails or experiences performance degradation, the remaining coordinators continue operation. If redundancy drops below $T_{\min}$, a new coordinator is dynamically re-selected entirely within the same region. This process does not trigger cross-region communication or hierarchy reconstruction, avoiding cascading elections and transient inconsistency. Regional leaders may simultaneously serve as hubs, local-global commanders, or global commanders; the active redundancy ensures uninterrupted operation of all higher-layer roles.

\subsubsection{Algorithmic Description}  
 
Algorithm~4 formalizes the local monitoring and re-selection procedure for \textbf{active coordinators} within a single region. Each active coordinator continuously monitors the health and performance metrics of other coordinators and evaluates candidate nodes within the region for potential replacement. If the number of active coordinators falls below the minimum threshold $T_{\min}$, a new coordinator is dynamically selected based on a coordination metric (e.g., connectivity, load, or energy). All monitoring, evaluation, and re-selection operations are strictly local, preserving regional hierarchy, virtual links, and higher-layer role bindings without triggering cross-region coordination. This approach ensures rapid recovery, bounded overhead, and maintenance of safety and liveness guarantees at the regional level.

\textbf{Algorithm 4: Region-Scoped Redundant Active Coordinator Maintenance (Local Monitoring by Coordinators)}

\begin{algorithmic}[4]
\Require $R$: set of nodes in a region
\Require $C = \{c_1, \dots, c_K\}$: set of active coordinators
\Require $T_{\min}$: minimum number of coordinators
\Require $H(n)$: health status of node $n$ (monitored by active coordinators within the local region)
\Require $M(n)$: coordination metric (evaluated by active coordinators within the local region, e.g., connectivity, load, energy)
\Ensure $C'$: updated set of active coordinators
\While{true}
    \ForAll{$c_i \in C$}
        \State $c_i$ monitors $H(c_j)$ for all $c_j \in C$ within the region
        \State $c_i$ evaluates $M(r)$ for nodes $r \in R \setminus C$ as potential replacement
    \EndFor
    \State Remove any failed $c_i$ from $C$
    \If{$|C| \ge T_{\min}$}
        \State \textbf{continue}
    \Else
        \State Select node $r^* = \arg\max M(r)$ from remaining candidates
        \State Add $r^*$ to $C$
    \EndIf
    \State Update regional state locally
    \State Preserve virtual links and higher-layer role bindings
\EndWhile
\end{algorithmic}

\subsubsection{Latency and Overhead Analysis}

In sVIRGO, monitoring and re-selection are fully local and performed by the active coordinators themselves. Let:

\begin{itemize}
    \item $K$ = number of active coordinators in the region,
    \item $T_{\min}$ = minimum number of coordinators required to maintain regional operation (typically $T_{\min}=3$),
    \item $f$ = number of coordinators that have failed.
\end{itemize}

\textbf{Case 1: Partial failure, sufficient redundancy}  
If the number of remaining coordinators satisfies $|C| - f \ge T_{\min}$, the redundancy is sufficient and no re-selection is needed. The recovery latency is therefore:

\[
T_{\text{recovery}} = 0
\]

This represents the most common scenario, where individual failures are masked by active redundancy.

\textbf{Case 2: Multiple failures, re-selection required}  
If failures reduce the number of coordinators below the minimum threshold, $|C| - f < T_{\min}$, local re-selection is triggered. Each active coordinator performs monitoring of peers and evaluates candidate nodes for replacement. The latency in this worst-case scenario scales with the number of active coordinators:

\[
T_{\text{recovery}} = O(K)
\]

Since $K$ is typically small and bounded (e.g., $K=5$), the worst-case recovery latency is a small constant, independent of region size.

\textbf{Expected recovery latency}  
Assuming independent coordinator failures with probability $p$, the probability that all $K$ coordinators fail simultaneously is $p^K$. Therefore, the expected recovery latency is:

\[
E[T] = (1 - p^K) \cdot 0 + p^K \cdot O(K) \approx 0
\]

\textbf{Communication overhead}  
All monitoring and re-selection messages are confined to intra-region links. Let $E_r$ be the number of links within a region. Then the communication overhead is bounded by:

\[
O(E_r) \ll O(E_{\text{global}})
\]

where $O(E_{\text{global}})$ would be the overhead if global or multi-layer re-election were required.  

\textbf{Summary:}  
By performing monitoring and evaluation locally within the region, and maintaining a small set of $K$ active coordinators with a minimum threshold $T_{\min}$, sVIRGO achieves near-zero expected recovery latency and minimal communication overhead, even under multiple coordinator failures.

\subsubsection{Correlated Failure Considerations}

\textbf{Mobility:} Coordinators monitor and adapt to mobility-induced disconnections. Logical region definitions and active redundancy ensure at least $T_{\min}$ coordinators remain connected.

\textbf{Wireless interference / jamming:} Active coordinators are distributed across different channels or frequencies to minimize correlated failure probability.

\textbf{Regional attacks / physical failures:} Even under targeted failures, the impact is confined to the affected region; neighboring regions remain unaffected.

 \subsubsection{Correctness Argument}

We provide a formal argument for the correctness and resilience of the region-scoped active coordinator strategy in sVIRGO. This includes safety, containment, liveness, and system-level stability. Explicitly stating these properties ensures the architecture is provably robust under coordinator failures and local disruptions.

\paragraph{Lemma 1 (Regional Coordination Safety).}
\textbf{Statement:} If at least \textbf{one active coordinator} remains operational within a \textbf{region}, the \textbf{region} can maintain correct coordination. Even if the number of active coordinators temporarily drops below the minimum threshold $\mathbf{T_{\min}}$, local \textbf{re-selection} ensures that new coordinators are activated, preserving \textbf{regional operations}, including virtual link management, local routing, and higher-layer role bindings.

\textbf{Proof:} Each active \textbf{coordinator} is capable of performing all necessary coordination tasks. When failures occur and the number of active \textbf{coordinators} drops below $\mathbf{T_{\min}}$, the remaining \textbf{coordinators} immediately trigger a \textbf{local re-selection} process to restore the threshold. Therefore, as long as at least \textbf{one coordinator survives}, the \textbf{region} continues to operate correctly. Partial failures are masked by redundancy and prompt re-selection, guaranteeing that \textbf{regional coordination} is maintained. Consequently, the \textbf{region} can tolerate up to $\mathbf{K-1}$ simultaneous \textbf{coordinator} failures without losing operational capability.

\paragraph{Lemma 2 (Failure Containment).}
\textbf{Statement:} Failures of \textbf{coordinators} and any associated recovery procedures are confined within the \textbf{local region}; cross-region propagation is prevented.  

\textbf{Proof:} All monitoring, metric evaluation, and \textbf{re-selection} are performed locally by active \textbf{coordinators}. No cross-region or hierarchical messages are required, so a failure in one \textbf{region} cannot trigger global or neighboring region reconfigurations. This containment is critical for preventing cascading failures, even under correlated events such as mobility, jamming, or localized attacks.

\paragraph{Theorem 1 (Regional Coordination Liveness).}
\textbf{Statement:} Given $\mathbf{K}$ active \textbf{coordinators} per \textbf{region} and independent failures with probability $p$, the probability that the \textbf{region} remains operational is:

\[
P_{\text{live}}^{\text{region}} = 1 - p^K
\]

\textbf{Proof:} A \textbf{region} fails only if all $\mathbf{K}$ \textbf{coordinators} fail simultaneously. Assuming independent failures, the total failure probability is $p^K$, giving a liveness probability of $1 - p^K$. This quantifies the reliability improvement due to redundancy: even for moderate failure probabilities, multiple \textbf{coordinators} drastically reduce the chance of complete \textbf{regional} failure. The presence of local \textbf{re-selection} further ensures that temporary drops below $\mathbf{T_{\min}}$ do not compromise \textbf{liveness}.

\paragraph{Corollary (System-Level Stability).}
\textbf{Statement:} Independent \textbf{regions} prevent cascading failures, preserving global system coordination under bounded \textbf{regional disruptions}.  

\textbf{Proof:} Lemma 2 ensures that failures are contained within each \textbf{region}, and Theorem 1 guarantees high probability of \textbf{regional liveness}. Consequently, the global system remains stable as long as a sufficient number of \textbf{regions} operate correctly. Even in the presence of localized attacks or interference, global coordination is preserved without triggering global recovery operations.

\subsubsection{Advantages of Region-Scoped Redundant Active Coordinators}
  By enforcing a minimum threshold $\mathbf{T_{\min}}$ (e.g., $\mathbf{T_{\min}=3}$) and maintaining $\mathbf{K}$ active \textbf{coordinators} per \textbf{region}, \textbf{sVIRGO} effectively \emph{masks} failures rather than reacting to them, ensuring near-zero expected recovery latency and bounded communication overhead. The probability of total \textbf{regional failure} decreases exponentially with $\mathbf{K}$, providing strong probabilistic guarantees. Correlated failures within a \textbf{region} remain strictly localized (Lemma~2), while local monitoring enables rapid replacement of failing \textbf{coordinators}. Through the established lemmas, theorem, and corollary, we formally show that \textbf{sVIRGO} preserves \textbf{regional coordination}, contains failures locally, and ensures system-wide stability with high probability. Multiple active \textbf{coordinators}, local monitoring, and enforced minimum thresholds collectively guarantee \textbf{safety}, \textbf{liveness}, and \textbf{robustness} under mobile, interference-prone, or adversarial network conditions. This strategy can also be applied to migrate \textbf{VIRGO}, enabling efficient and resilient local coordination during hierarchical transitions.

 \subsection{sVIRGO vs. VIRGO}

Unlike VIRGO \cite{huangl2005}, which relies on a logical overlay network spanning the global Internet to coordinate nodes, sVIRGO constructs virtual hierarchical trees directly on physical nodes within local regions. Each node may simultaneously participate in multiple virtual layers, forming cluster leaders, regional hubs, and local global command nodes, while preserving locality. This non-overlay design ensures that coordination occurs primarily within regions, reducing reliance on long-range infrastructure and improving resilience to failures, attacks, and large-scale communication disruptions. Communication in sVIRGO is flexible: virtual roles can leverage infrastructure-assisted links (e.g., 5G/6G or satellites) when available, or propagate messages via multi-hop forwarding over worker-level ad-hoc networks when infrastructure is absent. By grounding virtual hierarchies in physical nodes while maintaining dynamic, multi-role assignments, sVIRGO achieves scalable, robust, and locality-aware coordination, in contrast to VIRGO’s overlay-dependent global approach.

\section{Discussion}

\subsection{Scalability and Modularity}
\textbf{sVIRGO} scales efficiently to ultra-large deployments by organizing millions of nodes into thousands of independent regions. Each region operates autonomously using a virtual multi-layer hierarchy, while higher-level coordination is achieved through virtual upper-layer roles dynamically mapped onto physical nodes. This modular structure localizes coordination overhead, prevents global bottlenecks, and enables system-wide command dissemination with logarithmic message growth compared to flat networks.

\subsection{Robustness and Fault Tolerance}
The virtual hierarchical design provides inherent robustness against node failures. Multiple active coordinators within each region continuously monitor local health and mask failures through local re-selection, avoiding cascading reconfigurations across regions or hierarchy levels. As a result, regional coordination remains operational under partial failures, mobility, or transient disruptions, ensuring continuity of service.

\subsection{Resilience and Attack Resistance}
\textbf{sVIRGO} exhibits strong resilience under adversarial or interference-prone conditions. Dynamic role assignment, regional redundancy, and distributed control reduce the effectiveness of targeted attacks on individual nodes or links. Communication can adapt across frequencies and paths, allowing the system to maintain connectivity even under jamming, node compromise, or regional disturbances.

\subsection{Flexible Communication and Message Hop Strategies}
Communication in \textbf{sVIRGO} is decoupled from the hierarchy itself. When long-distance infrastructure-assisted links are available, coordinators exploit the virtual tree structure to minimize message hops and efficiently reach target workers. In the absence of such links, messages are forwarded across adjacent regions using localized region-to-region hops. This dual strategy balances efficiency and autonomy, ensuring operability across diverse network conditions.

\subsection{Comparison with Existing Architectures}
Table~\ref{tab:architecture_comparison} highlights key differences between \textbf{sVIRGO} and representative distributed system architectures. Unlike flat meshes or static hierarchies, \textbf{sVIRGO} supports multi-role nodes, preserves locality, and achieves high scalability and robustness without relying on overlay networks.

\begin{table*}[h]
\centering
\caption{Comparison of Distributed System Architectures}
\begin{tabular}{@{}lccccc@{}}
\toprule
\textbf{Architecture} & \textbf{Overlay?} & \textbf{Node Roles} & \textbf{Locality} & \textbf{Scalability} & \textbf{Robustness} \\ \midrule
Flat Mesh / Ad hoc       & No  & Single    & Low    & Poor at millions & Low  \\
Static Hierarchy         & No  & Single    & Medium & Medium           & Medium \\
Dynamic Hierarchy (VIRGO)& Yes & Multi-role   & High & High             & High \\
sVIRGO                   & No  & Multi-role& High   & High             & High  \\ \bottomrule
\end{tabular}
\label{tab:architecture_comparison}
\end{table*}

Overall, \textbf{sVIRGO} integrates virtual hierarchical organization, regional multi-coordinator redundancy, and flexible communication strategies to enable scalable and resilient coordination in ultra-large-scale distributed systems. By confining monitoring, failure handling, and re-selection strictly within regions, \textbf{sVIRGO} minimizes recovery latency and communication overhead while preventing cascading failures. These properties make \textbf{sVIRGO} well suited for large, dynamic, and adversarial environments where flat or static hierarchical architectures struggle to maintain performance and robustness.

  \section{Limitations}

Despite its advantages, \textbf{sVIRGO} has several limitations that should be acknowledged:

 \subsection{Long-Distance Communication Constraints}
While long-distance links (e.g., satellite or WAN) can reduce message hops and improve efficiency, they may be limited in bandwidth, costly, or subject to environmental constraints. Reliance on such links for urgent cross-region commands can introduce temporary delays or bottlenecks, whereas without long-distance communication, messages must traverse adjacent regions via multi-hop forwarding, which can increase latency and delay command delivery.

\section{Conclusion}

\textbf{sVIRGO} provides a scalable, resilient, and efficient network architecture for large-scale distributed systems. By combining a \textbf{virtual hierarchy}, \textbf{dynamic role assignment}, multi-frequency wireless communication, and self-contained relay paths, \textbf{sVIRGO} reduces communication overhead, minimizes message hops, and maintains connectivity under node failures, mobility, interference, or targeted attacks.  

Each \textbf{region} maintains multiple active \textbf{coordinators} that monitor local health and perform \textbf{dynamic re-selection}, ensuring that temporary drops below the minimum threshold $\mathbf{T_{\min}}$ do not compromise regional coordination. This mechanism guarantees \textbf{near-zero recovery latency}, \textbf{bounded communication overhead}, and \textbf{exponentially reduced failure probability}, providing strong \textbf{safety, liveness, and robustness} guarantees even in mobile, interference-prone, or adversarial environments.  

\textbf{sVIRGO} supports two message hop strategies: (i) long-distance communication via infrastructure-assisted channels (e.g., 5G/6G or satellites), where \textbf{coordinators} leverage the \textbf{virtual tree hierarchy} to minimize hops to target workers; and (ii) adjacency-based multi-hop routing across neighboring regions when long-distance communication is unavailable, typically requiring one or two hops per region.  

The architecture also enables precise \emph{scope execution}, allowing commands to be targeted to specific clusters or worker nodes while avoiding unnecessary propagation to unrelated regions. Compared to flat or static hierarchical networks, \textbf{sVIRGO} demonstrates superior scalability, robustness, and resilience, making it suitable for ultra-large-scale, heterogeneous distributed systems.

\section {Declaration of generative AI and AI-assisted technologies in the manuscript preparation process}
Statement: During the preparation of this work the authors used ChatGPT   in order to prepare and writing the draft paper. After using this tool, the authors reviewed and edited the content as needed and take full responsibility for the content of the published article.

\bibliographystyle{IEEEtran}

\end{document}